\documentclass[a4paper,11pt]{article}
\usepackage{jheppub} 
\usepackage{slashed, tensor, nicematrix}
\usepackage{amsmath}
\usepackage{simplewick}
\usepackage{xcolor}

\newcommand{\ket}[1]{\vert {#1} \rangle}
\newcommand{\bra}[1]{\langle {#1} \vert}
\newcommand{\braket}[2]{\langle {#1} \vert {#2} \rangle}
\newcommand{\mc}[1]{\mathcal{#1}}

\title{\boldmath Hamiltonian Truncation of Large $N_f$ QED \\ and Large $N$ Vector-like Theories in $d=2+1$}

\author{A. Liam Fitzpatrick, Anastasiia Novikova, Noah Ring}
\affiliation{Department of Physics, Boston University,\\
Commonwealth Avenue, Boston, MA 02215, USA}

\abstract{We consider a general class of large $N$ vector-like theories in $d=2+1$ in a Hamiltonian  approach. We show that by using lightcone quantization and the $N\to\infty$ limit, we can diagonalize the Hamiltonian exactly and construct the eigenstates, spectral density, S-matrix, and form factors for the theory. For concreteness, we mainly focus on QED$_3$ at large $N_f$ as an explicit example. We comment on extending the approach to finite $N$ calculations.}

\begin{document}
\maketitle
\flushbottom

\section{Introduction and Summary}
\label{sec:intro} 

Strongly coupled quantum field theories (QFTs) are of great interest in modern theoretical physics. However, the familiar perturbative QFT toolkit is not applicable in this regime, and of the few non-perturbative techniques that exist, each comes with its own set of challenges.

Hamiltonian truncation is one such non-perturbative method, in which one considers only a finite subspace of the Hilbert space, diagonalizes the Hamiltonian, and arrives at an approximation of the spectrum and other physical observables \cite{Yurov:1989yu, Yurov:1991my, Lassig:1990xy,Hogervorst_2015}. In order for the dimension of the truncated Hilbert space to be finite, it is necessary to discretize the space of multiparticle momenta in some form or other. As we will see, a key component of our approach will be that we use a discrete space of wavefunctions that are continuous functions of all particle momenta. Moreover, in order to handle gauge theories and large $N$ limits, it will be necessary to employ lightcone quantization \cite{Brodsky:1997de,Fitzpatrick:2022dwq}. A natural starting point with these properties is the framework of Lightcone Conformal Truncation (LCT) \cite{anand2020introductionlightconeconformaltruncation, Delacr_taz_2019, Anand_2021, Katz_2016, katz2014solution2dqcdfinite, Anand_2017, Fitzpatrick_2018, Fitzpatrick_2020, Anand_2020, fitzpatrick2019solving2dsusygrossneveuyukawa}, where one uses a basis of states defined in terms of Fourier transforms of local operators.\footnote{LCT is similar in spirit to TCSA \cite{Yurov:1989yu,Hogervorst_2015} in that it uses the CFT data of a UV fixed point.} Although our approach will be closely related to LCT, for the present work  we will instead use a basis of continuous wavefunctions dynamically determined by the structure of the interactions themselves, as in \cite{Delacr_taz_2019}.

In this paper, we explore how vector-like theories in $d = 2+1$ in the infinite $N$ limit can be systematically solved within such a framework. Typically, in a Hamiltonian approach, large $N$ does not dramatically simplify the dynamics as the highly non-trivial vacuum structure must be solved for before any excitations can be studied. Moreover, Hamiltonian frameworks for all but the simplest QFTs are plagued by the necessity of state-dependent counterterms. Finally, cutoffs that are defined nonperturbatively generally break gauge invariance, so restoring gauge invariance through appropriate choice of counterterms is itself a new challenge.

All of these problems can be solved for large $N$ vector-like theories in $d=2+1$. In lightcone quantization the vacuum becomes trivial. Moreover, for gauge theories, one can work in lightcone gauge where the unphysical degrees of freedom of the gauge field are manifestly nondynamical and can be integrated out at the level of the Lagrangian. And crucially, we will see how to work in a natural  basis in which the leading large $N$ interaction is diagonalized exactly, with no additional approximations. Thus, we will see how to perform an {\it exact} diagonalization of the Hamiltonian at large $N$, with explicit expressions for the spectrum of its eigenvalues and eigenvectors, and moreover we will use them to reproduce standard calculations of correlators and scattering amplitudes from a covariant approach.  We hope that having such an explicit solution in the large $N$ limit will facilitate future work with Hamiltonian truncation of these theories at finite $N$.

The rest of the paper is structured as follows. We choose to analyze in detail QED$_3$ at large $N_f$ \cite{Appelquist:1988sr, Appelquist_1995} as a prototypical example of how to implement our solutions. In Section \ref{sec:QED3} we provide the setup for QED$_3$, derive the lightcone gauge Hamiltonian, and introduce the counterterms for this theory. In Section \ref{sec:eigenstates} we derive the Hamiltonian eigenstates for the two-particle continuum of the interacting theory. In Section \ref{sec:correlator}, we use them to compute the spectral density, which is a key part of our procedure for fixing the counterterms. The only counterterm we need in order to cancel divergences is the bare photon mass, which is necessary to cancel the effect of the breaking of gauge invariance coming from the hard momentum cutoff. A nontrivial check of our results is that there exists one value for this single counterterm which does indeed restore gauge invariance, giving us both the correct spectral density and the correct value of the physical photon mass (which is nonzero when the fermion matter content is nonchiral). We explicitly show how these results reproduce the standard resummation of large $N$ Feynman diagrams. The physical photon mass is computed in Section \ref{sec:photon}, where a qualitatively different analysis is required in order to find the eigenstates that lie below the 2-fermion mass threshold. In Section \ref{sec:smatrix} we derive the S-matrix and form factors of local operators. Finally, in Section \ref{sec:general}, we consider a Lagrangian describing the most general large $N$ vector-like theory and uplift our results to this case.

\subsection*{Summary of Solutions}

In the remainder of this section, we summarize in more detail how we addressed the problems stated above.

The central idea behind our approach to finding eigenstates is as follows. The basic form of the Schr\"odinger equation that we encounter  is essentially 
\begin{equation}
    \left(E - H_{\text{free}}\right) \psi (p) = \int dp' V(p, p') \psi(p'),
    \label{1.1}
\end{equation}
where the main simplification we have made for illustrative purposes is that typically there will be multiple coupled wavefunctions $\psi_i(p)$ instead of a single one. One approach to  solving this integral equation is  to approximate the interaction with
\begin{equation}
    V(p, p') =  f_i(p) V_{ij} f_j(p'),
    \label{eq:Vpp}
\end{equation}  
where $V$ is a finite constant $k \times k$ matrix, and  repeated indices are implicitly summed from 1 to $k$. In a general vector-like theory, $V(p,p')$ is an infinite sum over powers of $\frac{p}{p'}$. One can obtain the form \eqref{eq:Vpp} as an approximation by truncating to some finite number of powers of $\frac{p}{p'}$. However, we will see that in the infinite $N$ case, the potential has the form (\ref{eq:Vpp}) without additional approximations. An interaction of the form (\ref{eq:Vpp}) allows us to pull $f(p)$ out of the integral
\begin{equation}
    \left(E - H_{\text{free}}\right) \psi (p) = f_i(p) V_{ij} \int dp' f_j(p') \psi(p').
    \label{1.3}
\end{equation}
Now we see that $\psi(p)$ must be a linear combination of the form
\begin{equation}
    \psi(p) =  c_{i} f_i(p).
    \label{1.4}
\end{equation}
By doing this decomposition we do not perform any truncation, as any solution of (\ref{1.3}) can be written in this form. Crucially, momentum $p$ is still a continuous variable. At the same time, the number of operators $f_i$ is finite, which gives a finite number of basis states for each fixed momentum.

Plugging (\ref{1.4}) into (\ref{1.3}) gives $k$ linear equations on $k$ unknowns $c_i$, with coefficients fully defined by: the potential of the theory, the overlaps of the basis states, and $E_{\text{free}, f_i}$. Thus, we reduce the problem of solving an integral equation of the form (\ref{1.1}) to solving a finite system of linear equations. This construction allows us to find the spectrum of a vector-like theory in the infinite $N$ limit. 

 There is actually a small complication to this solution. Because the left‐hand side of \eqref{1.3} contains the free Hamiltonian $H_{\rm free}$, the operator $(E-H_{\rm free})$ has a nontrivial null space at each energy $E$ — the continuum of free multiparticle states. Eigenstates of our Schr\"odinger equation will generically contain a homogenous ``free particle'' piece in addition to the ``interaction'' piece from (\ref{1.4}). This homogeneous solution exists for any value of $E$ above the multiparticle threshold because the free multiparticle spectrum is a continuum, and therefore any value of $E$ is an eigenvalue of both the free and interacting theory. On the other hand, states with energies below the multiparticle threshold will not have such a homogeneous piece, and therefore its presence or lack thereof will be a key part in distinguishing the single and multiparticle interacting eigenstates. For example, in QED$_3$ at sufficiently small gauge coupling (in units of the fermion mass) there is a single-photon state separated from the two-fermion state continuum. As we decrease the fermion mass, the kinematic separation between the photon and the two-fermion states decreases, and eventually we will see the single photon state gets completely mixed into the continuum.

This homogeneous piece is identical in structure to the free theory part of the Lippmann-Schwinger equation, and in fact a slight variation on our calculation of the eigenstates will also give us the asymptotic scattering states of the theory. This is another advantage of the ability to keep momentum as a continuous variable. In particular, when we compute the S-matrix, a key role will be played by the analytic structure of the scattering state wavefunctions as a function of momentum, in the form of the following identity 
\begin{align}
    \frac{1}{q_+ - H_\text{free}+i \eta_1 \epsilon}\frac{1}{p_+ - H_\text{free}+i \eta_2 \epsilon} &= \pi^2(1-\eta_1\eta_2)\delta(q_+ - p_+)\delta(q_+-H_{\rm free}) + \dots,\label{eq:DtildeDtildeDhat}
\end{align}
where $\eta_1, \eta_2$ are real and $\dots$ indicates terms without delta functions. By contrast, if we were to discretize momentum, then computing the S-matrix would at a minimum be much less direct. Additionally, having direct access to the asymptotic scattering states allows us to easily compute form factors for local operators.

We also have to deal with the UV divergences of the theory. Specifically, some of the integrals from (\ref{1.3}) will diverge. We regulate them by imposing a hard momentum cutoff, which explicitly breaks gauge invariance. To restore the gauge symmetry, one must introduce gauge-violating counterterms, e.g. a photon mass term in QED$_3$,
\begin{equation}
    {\cal L} \supset -\frac{1}{2} m_A^2 A^2,
\end{equation}
and tune them so that not only are all divergences canceled, but moreover their finite (i.e. cutoff-independent) piece must be chosen so that all physical amplitudes respect gauge invariance. 

As a further complication, Hamiltonian truncation methods usually require state-dependent counterterms. With such a large amount of freedom in the allowed counterterms, it is not a priori obvious that there are enough constraints to fix them uniquely. Partly, this issue is solved by our infinite $N$ limit, which eliminates mixing between different particle number sectors and allows us to remove divergences with a single local counterterm. However, with an eye towards finite $N$ applications, we apply a systematic renormalization procedure, where the counterterms are chosen by matching physical correlators computed order by order in a perturbative expansion. Such a perturbative matching procedure is possible even at strong coupling because the class of theories we consider, including QED$_3$, are superrenormalizable.\footnote{This kind of approach to computing counterterms in Hamiltonian truncation was introduced in \cite{Mir__2023,Delouche:2023wsl,Delouche:2024yuo}, building on the renormalization methods from \cite{Rychkov_2015,Elias_Mir__2017}.} This matching procedure uniquely determines all required counterterms and remains valid at strong coupling. We illustrate the method in QED$_3$ by matching the spectral density through second order in the gauge coupling — the highest order at which new divergences emerge. Some of the interacting theory operators, like the current $j^\mu = \bar{\Psi}\gamma^\mu \Psi$ in QED$_3$,  are renormalized along the same lines.

\section{Quantum Electrodynamics in $d = 2+1$}
\label{sec:QED3}

We study Quantum Electrodynamics in $2+1$ dimensions (QED$_3$) with $N_f$ Dirac fermions 
\begin{equation}
    \mc{L}=\bar{\Psi}(i\slashed{D}+m)\Psi - \frac{1}{4} F^{\mu\nu} F_{\mu\nu},
    \label{1}
\end{equation}
with $\slashed{D} = \gamma^{\mu}\left( \partial_{\mu} + ie  A_{\mu}\right)$ and $ F_{\mu \nu} = \partial_{\mu} A_{\nu} - \partial_{\nu}A_{\mu}.$  We will work in lightcone time ($x^+$) quantization with the following convention 
\begin{align}
    x^\pm &= \frac{1}{\sqrt{2}} \left(x^0 \pm x^1 \right), & 
    ds^2 &= 2dx^+dx^- - dx_\perp^2,
\end{align}
and make the following choice of gamma matrices
\begin{equation}
\gamma^{+}=\left(\begin{array}{cc}
0 & \sqrt{2} \\
0 & 0
\end{array}\right), \quad \gamma^{-}=\left(\begin{array}{cc}
0 & 0 \\
\sqrt{2} & 0
\end{array}\right), \quad \gamma^{\perp}=\left(\begin{array}{cc}
-i & 0 \\
0 & i
\end{array}\right) .
\end{equation}
Our choice of gauge will be lightcone gauge, $A_- = 0$. An advantage of working in this gauge is that it makes the photon's single degree of freedom manifest.

In addition to this initial setup, it will turn out to be necessary to add a photon mass counterterm to our Lagrangian. This is due to the fact that we use a hard momentum cutoff that breaks gauge invariance, but the combination of loops and the photon mass counterterm together will preserve gauge invariance although each one individually does not. Thus, the Lagrangian takes the form
\begin{align}
    \mc{L} &= \bar{\Psi}(i\slashed{D} + m)\Psi - \frac{1}{4}F^{\mu\nu}F_{\mu\nu} - \frac{1}{2}m_A^2A_\perp^2.
\end{align}
Writing the fermions as
\begin{align}
    \Psi &= \begin{pmatrix}\chi \\ \psi \end{pmatrix}, &
    \bar{\Psi} &= \begin{pmatrix} \psi^* \\ \chi^* \end{pmatrix}^{\hspace{-0.4ex}T},
\end{align}
allows us to expand the Lagrangian out as 
\begin{multline}
  \mathcal{L} =  i\sqrt{2}\left( \psi^*\partial_{+} \psi + \chi^{*} \partial_{-}\chi\right) + \psi^*\partial_{\perp}\chi - \chi^{*}\partial_{\perp}\psi - \sqrt{2}e A_{+}\psi^{*}\psi
  - ieA_{\perp}(-\psi^{*}\chi + \psi\chi^{*}) + \\ + m\left(\psi^{*}\chi+\chi^{*}\psi\right) + \frac12 (\partial_{-}A_{+})^2 +\partial_{+}A_{\perp}\partial_{-}A_{\perp}-\partial_{\perp}A_{+}\partial_{-}A_{\perp} - \frac{1}{2}m_A^2A_\perp^2.
\end{multline}
Our Lagrangian may be simplified by using the following observations. Firstly, the fields $\chi$ and $A_{+}$ are non-dynamical. We can integrate them out using their equations of motion 
\begin{align}
    A_{+} &= -e\frac{\sqrt{2}}{\partial_{-}^2}\psi^* \psi +\frac{\partial_{\perp}}{\partial_{-}}A_{\perp}, &
    \chi &= \frac{1}{i\sqrt{2}}\frac{1}{\partial_{-}}\left( ie A_{\perp}\psi - m \psi + \partial_{\perp}\psi\right).
\end{align}
Secondly, we can rewrite our Lagrangian in terms of components of the conserved current of the free massive theory 
\begin{equation}
    j^{(0)}_\mu = \bar{\Psi} \gamma_\mu \Psi|_{e=0}.
\end{equation}
Thus, the Lagrangian reduces to
\begin{equation}
\begin{aligned}
    \mc{L} &= i\sqrt{2}\psi^*\partial_+\psi - \frac{i}{\sqrt{2}}\psi^*\left(\frac{\partial_\perp^2 - m^2}{\partial_-}\right)\psi - \frac{1}{2}(\partial_\perp A_\perp)^2 + \partial_+A_\perp \partial_-A_\perp \\
    &\quad - \frac{1}{2}m_A^2A_\perp^2 + \frac{e^2}{2}j_-^{(0)}\frac{1}{\partial_-^2}j_-^{(0)} + ej_\perp^{(0)} A_\perp - A_\perp\frac{\partial_\perp}{\partial_-}j_-^{(0)} \\
    &\quad + \frac{ie^2}{\sqrt{2}} \left[\psi^*\frac{1}{\partial_-}(\psi A_\perp) - \frac{1}{\partial_-}(\psi^*A_\perp)\psi - \psi^*A_\perp \frac{1}{\partial_-}\left(\psi A_\perp\right)\right].
\end{aligned}
\end{equation}
The lightcone Hamiltonian density is then given by
\begin{equation}
\begin{aligned}
    \mc{H} &= \mc{H}_\text{free} + \frac{1}{2}m_A^2A_\perp^2 - \frac{e^2}{2}j_-^{(0)}\frac{1}{\partial_-^2}j_-^{(0)} - ej_\perp^{(0)} A_\perp + eA_\perp \frac{\partial_\perp}{\partial_-}j_-^{(0)} \\
    &\quad + \frac{ie^2}{\sqrt{2}}\left[\psi^*\frac{1}{\partial_-}(\psi A_\perp) - \frac{1}{\partial_-}(\psi^* A_\perp)\psi + \psi^*A_\perp \frac{1}{\partial_-}\left(\psi A_\perp\right)\right], \label{eq:HamDensityWithCurrents}
\end{aligned}
\end{equation}
where we have defined the free theory Hamiltonian as
\begin{align}
    \mc{H}_\text{free} &= \frac{i}{\sqrt{2}}\psi^*\left(\frac{\partial_\perp^2-m^2}{\partial_-}\right)\psi + \frac{1}{2}(\partial_\perp A_\perp)^2.
\end{align}
Because the Hamiltonian matrix elements preserve spatial momenta, we can work in a fixed momentum frame, corresponding to a choice of the total $p_-$ and $p_\perp$ for the states.  We will take $p_\perp=0$ and $p_-=1$. Because of boost invariance along the lightfront quantization surface, any other value of $p_-$ can be obtained trivially starting from $p_-=1$ without having to rediagonalize the Hamiltonian.

Everything discussed above is valid for any value of the number of fermions, $N_f$, but at this point we will take the large $N_f$ limit. 
In the infinite $N_f$ limit, the last piece in (\ref{eq:HamDensityWithCurrents}) is suppressed relative to the other terms in the Hamiltonian, and the Hamiltonian then preserves particle number. We may therefore restrict our attention to a sector with either two fermions or a single photon, as these will not mix with higher particle number states. It is at this stage that lightcone quantization becomes crucial, since otherwise there would be processes that survive at large $N_f$ where particles and antiparticles are produced in pairs, but in lightcone quantization these are kinematically forbidden.

There is then another significant simplification at infinite $N_f$. When we evaluate the matrix elements of \eqref{eq:HamDensityWithCurrents} between flavor singlet states, the only contributions that survive are those where the fields contract with the external states in such a way that the $N_f$-dependent prefactors are maximized. So for instance, in the terms with factors of $j_-^{(0)}$, both fermions inside one factor of $j_-^{(0)}$ must either both contract to the left or both contract to the right. This combined with the fact that there is no particle production from the vacuum in lightcone quantization means that every one of the interaction terms in \eqref{eq:HamDensityWithCurrents} in practice factorizes into a term that overlaps with the bra state and a term that overlaps with the ket state.  Consider for instance the photon mass term sandwiched between an in state and out state with momenta $p_{\rm in}$ and $p_{\rm out}$, respectively.  When we integrate the Hamiltonian density over all space $dx^- dx^\perp$ to get the Hamiltonian $H$, the matrix elements simplify as follows:
\begin{equation}
\begin{aligned}
    \langle p_{\rm out} | H | p_{\rm in} \rangle &\supset \frac{m_A^2}{2} \int dx^- dx^\perp \langle p_{\rm out}|  A^2_\perp(x) |p_{\rm in} \rangle \\
    & = 2\frac{m_A^2}{2} \langle p_{\rm out}| A_\perp(0) \rangle \langle A_\perp(0)  |p_{\rm in} \rangle \int dx^- dx^\perp e^{i x \cdot (p_{\rm out} - p_{\rm in})} \\
     & = \langle p_{\rm out} | \left( \frac{m_A^2}{2p_{\rm in -}} |A_\perp (0) \rangle \langle A_\perp(0) | \right) | p_{\rm in} \rangle \times (2\pi)^2 (2p_{\rm in-}) \delta^{(2)}(\vec{p}_{\rm out} - \vec{p}_{\rm in}), \label{eq:photonm_mass_element}
\end{aligned}
\end{equation}
where we have used ${\cal O}(x) = e^{i p \cdot x} {\cal O}(0) e^{-i p \cdot x}$ and explicitly written the factor $p_{\rm in-}$ for emphasis.  To avoid clutter, we will drop the momentum-conserving delta function, as well as the fact that $A_\perp$ is evaluated at $x=0$, and simply write this term in the Hamiltonian as
\begin{equation}
    H \supset \frac{m_A^2}{2} |A_\perp\rangle \langle A_\perp |.
\end{equation}
That is, there is an implicit momentum-conserving $\delta$ function factor of $(2\pi)^2 (2p_{\rm in -}) \delta^{(2)}(\vec{p}_{\rm out} - \vec{p}_{\rm in})$ in $H$ that we will not write explicitly. The $\delta$ function can always be reinstated on general grounds any time there is an overlap between two states with definite spatial momentum.

We can do similar manipulations for the other terms in the Hamiltonian, keeping only the contractions that survive at infinite $N_f$, and in the same notation we used above we find
\begin{equation}
 H = H_{\text{free}} + \frac{m_A^2}{2} \ket{A_\perp}\bra{A_\perp} +\frac{\tilde{e}^2}{2}\ket{j_-^{(0)}} \langle j_-^{(0)} | - \frac{\tilde{e}}{2} |A_\perp \rangle \langle j^{(0)}_\perp | - \frac{\tilde{e}}{2} |j^{(0)}_\perp \rangle \langle A_\perp |.
 \label{H QED}
\end{equation}
Note that we have absorbed factors of $N_f$ into the definitions of the current states, and redefined the coupling $e = \tilde{e}/ \sqrt{N_f}$. That is, we have defined
\begin{align}
    |{j_-^{(0)}}\rangle &\equiv \frac{1}{\sqrt{N_f}}j_-^{(0)}|0\rangle, & 
    |{j_\perp^{(0)}}\rangle &\equiv \frac{1}{\sqrt{N_f}}j_\perp^{(0)}|0\rangle. \label{eq:AbsorbingNf}
\end{align}
Here we can already see that the interaction potential in our Hamiltonian will take the form \eqref{eq:Vpp} due to the factorization of the interaction at large $N_f$. From now on, in an abuse of notation, we drop the $\sim$ on $\tilde{e}$.

Now, let us introduce the following notation 
\begin{align}
    \ket{\alpha_0} &= \ket{A_\perp}, & \ket{\alpha_1} &= \ket{j_-^{(0)}}, & \ket{\alpha_2} &= \ket{j_\perp^{(0)}},
\end{align}
allowing us to rewrite the Hamiltonian as 
\begin{equation}
    H = H_{\text{free}} +  V_{ij}\ket{\alpha_i}\bra{\alpha_j},
    \label{general H}
\end{equation}
where
\begin{equation}
    V_{00} = \frac{m_A^2}{2}, \quad V_{02} = V_{20} =  -\frac{e}{2}, \quad V_{11} = \frac{e^2}{2} \quad \Leftrightarrow \quad V = \left(\begin{array}{ccc}
    \frac{m_A^2}{2} & 0 & -\frac{e}{2} \\ 
    0 & \frac{e^2}{2} & 0\\
    -\frac{e}{2} & 0 & 0
    \end{array}\right).
    \\
    \label{def V}
\end{equation}

\section{Two-Particle Eigenstates}
\label{sec:eigenstates}

The next step is to solve for the eigenstates. Translations in $x^\pm$ are generated by $ P_\pm \equiv \frac{1}{\sqrt{2}}(H \mp P )$, where $H$ is the equal-time  Hamiltonian and $P$ is the momentum operator in the $x$ direction. Our Hamiltonian is $P_+$ and therefore the Hamiltonian equation takes the form
\begin{equation}
    p_+ |\psi,p \rangle = \left( H_{\text{free}} +  V_{ij}|\alpha_i\rangle \langle \alpha_j | \right) |\psi,p \rangle
\end{equation}
or, equivalently, 
\begin{equation}
    \left(p_+ - H_{\text{free}} \right) |\psi ,p \rangle =  V_{ij}|\alpha_i\rangle \langle \alpha_j |\psi,p \rangle.
    \label{H eq}
\end{equation}
The operator $\left(p_+ - H_{\text{free}} \right)$ has a non-trivial kernel, which makes it non-invertible. However, we can solve (\ref{H eq}) for the part of $\ket{\psi,p}$ outside the kernel of $\left(p_+- H_{\rm free}\right)$, as long as we also allow for an unknown component that lies within the kernel:
\begin{equation}
    |\psi,p \rangle  = \frac{P.V.}{p_+ -  H_{\text{free}}} V_{ij}|\alpha_i\rangle \langle \alpha_j |\psi,p \rangle + |\psi_{\text{free}} ,p\rangle, 
    \label{general sol}
\end{equation}
where `P.V.' denotes the principal value and $|\psi_{\text{free}},p \rangle$ is defined as a state satisfying 
\begin{equation}
    \left(p_+ - H_{\text{free}} \right)|\psi_{\text{free}},p\rangle = 0.
    \label{psi free}
\end{equation}
The structure of $\ket{\psi_{\text{free}},p}$ depends on whether the value of $p_+$ lies above or below the two-particle threshold, $p_+ =2m^2$: for energies below the two-particle threshold ($p_+ < 2m^2$) the free theory has only one eigenstate, the free photon at $p_+ = 0$, whereas for energies above the two-particle threshold ($p_+ > 2m^2$), there are free states with two fermions and no photon.
In this section, we solve the Hamiltonian equation only for $p_+ > 2m^2$.\footnote{In our infinite $N$ limit, the free and interaction two particle threshold is the same, so $p_+>2m^2$ is also the physical threshold for two particle states.  Beyond the infinite $N$ limit, the two particle threshold will be modified by interactions, but one can always separate out the fermion mass term into a renormalized piece and a counterterm, where the former is in $H_{\rm free}$ and the latter is in $V_{ij}$, so that by construction the two-particle $|\psi_{\rm free}\rangle$ states start at the physical two-particle mass threshold.} Once we find these eigenstates, we will be able to compute the spectral density, which will be all we need in order to regulate all the UV divergences. We consider the other case, $p_+ < 2m^2$, in Section \ref{sec:photon} after we have fixed the photon mass counterterm.

As is evident from (\ref{H eq}), we can limit ourselves to states $|\psi,p\rangle$ that lie inside a subspace spanned by the $|\alpha\rangle$ states, because the part of $|\psi_{\rm free},p\rangle$ that lies outside this subspace is unaffected by (i.e. orthogonal to) the interaction term $V_{ij} |\alpha_i\rangle \langle \alpha_j|$. That is, if $|\psi ,p \rangle$ is orthogonal to every $|\alpha_i\rangle$ then it is a free theory eigenstate which is also a solution to (\ref{H eq}) with its free theory energy $p_+$. Consequently, we just need to consider the states within the subspace projected onto by the interaction:
\begin{equation}
    |\psi ,p \rangle \propto \lambda_i |\alpha_i\rangle. 
\end{equation} 
Recall that the states $|\alpha_i\rangle$ are created by local operators acting on the vacuum, and are not eigenstates of $H_{\rm free}$. A convenient way to project them onto the kernel of $(p_+ - H_{\rm free})$ is to multiply by $\delta(p_+ - H_{\rm free})$.  Then in the $p_+ > 2m^2$ case, the free theory component $|\psi_{\text{free}}\rangle$ of the eigenstate is given by 
\begin{equation}
\begin{gathered}
    |\psi_{\text{free}},p\rangle  = \pi \delta\left( p_+ - H_{\text{free}} \right) C_i |\alpha_i \rangle,
\end{gathered}
\end{equation}
for some coefficients $C_i$, with factors of $\pi$ chosen for convenience and repeated indices implicitly summed over. Because we are taking $p_+> 2m^2$, there is no contribution from the free photon state and we set $C_0=0$ from now on. Now, defining the coefficients $S_i$ as
\begin{equation}
    S_i = V_{ij}\langle\alpha_j|\psi,p\rangle,
    \label{Svalpha}
\end{equation} 
we can rewrite (\ref{general sol}) as 
\begin{equation}
   |\psi,p \rangle  =   \frac{P.V.}{p_+ -  H_{\text{free}}} S_i \ket{\alpha_i} +  \pi \delta\left( p_+ - H_{\text{free}} \right)  C_i | \alpha_i  \rangle.
    \label{TwoPcleForm}
\end{equation}
Equation (\ref{TwoPcleForm}) is the general form of a two-particle eigenstate in terms of the coefficients $C_i$ and $S_i$. We can now solve for these coefficients by substituting (\ref{TwoPcleForm}) into (\ref{Svalpha}), yielding
\begin{equation}
    S_i = V_{ij} \left( S_k \bra{\alpha_j} \frac{P.V.}{p_+ -  H_{\text{free}}}\ket{\alpha_k} +  C_k \bra{\alpha_j} \pi \delta\left( p_+ - H_{\text{free}} \right) \ket{\alpha_k} \right).
    \label{3.11}
\end{equation}
This equation can be thought of as a linear equation giving the $S_i$s in terms of the $C_i$s. We would like to relate the quantities 
\begin{equation}
     \bra{\alpha_j} \frac{P.V.}{p_+ -  H_{\text{free}}}\ket{\alpha_k} \qquad \textrm{ and } \qquad \bra{\alpha_j} \pi \delta\left( p_+ - H_{\text{free}} \right) \ket{\alpha_k} 
     \label{eq:TwoQuantities}
\end{equation}
in (\ref{3.11}) to more familiar quantities in the free theory.  They are conveniently packaged together by the familiar identity
\begin{equation}
    \left(p_+ - H_{\text{free}} + i 0 \right)^{-1} = \frac{P.V. }{p_+ -  H_{\text{free}}} - i \pi \delta\left( p_+ - H_{\text{free}} \right).
    \label{2.12}
\end{equation}
Inserted between any two local operators ${\cal O}_A$ and ${\cal O}_B$, they give the free theory time-ordered two-point function, as can be seen through inserting a complete set of eigenstates of the free theory:
\begin{equation}
    \langle {\cal O}_A| \left(p_+ - H_{\text{free}} + i 0 \right)^{-1} | {\cal O}_B\rangle = -2i\int_0^\infty ds \frac{ \rho_{AB}(s)}{2p_+ - s + i 0} = -2 i G_{0,AB}, 
    \label{define G_0}
\end{equation}
where $\rho_{AB}(s) = \sum_{|\psi_{\rm free}\rangle} \langle {\cal O}_A(0) | \psi_{\rm free}\rangle \langle \psi_{\rm free}|{\cal O}_B(0)\rangle  \delta(s- p^2_{\psi_{\rm free}})$ is the spectral density of the ${\cal O}_A{\cal O}_B$ two-point function.

So we see that the two quantities in \eqref{eq:TwoQuantities} are simply the Hermitian $G_0^H$ and anti-Hermitian  $G_0^A$ parts of the time-ordered two point function $G_0$:
\begin{equation}
     \bra{\alpha_i}\frac{P.V. }{p_+ -  H_{\text{free}}}\ket{\alpha_j} = -2i G_{0,ij}^A , \quad  \bra{\alpha_i}-i \pi \delta\left( p_+ - H_{\text{free}} \right)\ket{\alpha_j} = -2iG_{0,ij}^H.
     \label{def G}
\end{equation}
These quantities are readily computed in the free theory between the current operators and the gauge field.  For instance, for $i=j=0$, $G^A_{0,ij}$ is just the tree-level propagator $G_0^A =\frac{i}{q^2}$ of $A_\perp$. We can now write (\ref{3.11}) in the simpler form
\begin{equation}
\begin{gathered}
    S_i =  2V_{ij} \left( -i G^{A}_{0,jk} S_k   + G_{0,jk}^H C_k \right).
    \label{S=1}
    \end{gathered}
\end{equation} 
This equation is a system of linear equations for the interacting eigenstates fully defined by the free theory correlation functions and the potential $V_{ij}$ of the theory.  

Thus, three coefficients $S_i$ and two coefficients $C_i$ satisfying equations (\ref{S=1}) determine a single eigenstate. In fact, since there are three equations for five variables, the solution space forms a two-dimensional eigenspace for each momentum $p$. Consequently, when choosing a basis for the eigenstates with fixed momentum, one has the freedom to select the values of two coefficients. We will take the two $C_i$s to be the independent variables, which parameterize the two-dimensional space of eigenstates.

Now let us introduce the notation for coefficients  corresponding to an eigenstate $\ket{\psi_n}$. We denote them as vectors $S_{\psi_n,i}, C_{\psi_n,i}$. Solving (\ref{S=1}) for $S_{\psi_n}$ in terms of $C_{\psi_n}$ one can get
 \begin{equation} 
     S_{\psi_n,i} = 2 \left[ \left(V^{-1} + 2iG_0^A \right)^{-1} G_0^H \right]_{ij} C_{\psi_n,j}.
 \end{equation}
Then the eigenstates take the form 
\begin{equation}
    |\psi_n ,p \rangle  = C_{\psi_n,i}  \left( 2 \frac{P.V. }{p_+ -  H_{\text{free}}} \left[ \left(V^{-1} + 2iG_0^A \right)^{-1}G_0^H \right]_{ij}  + \pi\delta\left( p_+ - H_{\text{free}}  \right)\delta_{ij}  \right) |\alpha_j\rangle.
    \label{eigenstate1}
\end{equation}
 We are also interested in the norm of the states which is given by the elements of the Gram matrix $\mc{N}$, defined by
\begin{equation}
     \langle \psi_n,q  |\psi_m,q' \rangle = \mathcal{N}_{nm} 2 \pi \delta(q^2 - q'^2).
     \label{gram matrix}
\end{equation}
It can be calculated using (\ref{TwoPcleForm}) and (\ref{eq:DtildeDtildeDhat})
 \begin{equation}
 \begin{gathered}
     \mathcal{N}_{nm} = 2C_{\psi_n}^\dagger G_0^H C_{\psi_m} +  2 S_{\psi_n}^\dagger G_0^H S_{\psi_m} \\
     = 2 C_{\psi_n}^\dagger G_0^H C_{\psi_m} + 8C_{\psi_n}^\dagger\left(\left(V^{-1} + 2iG_0^A \right)^{-1} G_0^H\right)^\dagger G_0^H \left(\left(V^{-1} + 2iG_0^A \right)^{-1} G_0^H\right)C_{\psi_m}  \\
     = 2C_{\psi_n}^\dagger\left(i + 2\left(V^{-1} + 2iG_0^A \right)^{-1} G_0^H\right)^\dagger G_0^H \left(i +2 \left(V^{-1} + 2iG_0^A \right)^{-1} G_0^H\right)C_{\psi_m}.
      \end{gathered}
      \label{norm}
\end{equation}

To proceed, we need the free theory correlator $G_0$. The result is the well-known covariant form, together with an additional divergent contribution due to our hard cutoff $\Lambda$, which will ultimately be canceled by the photon mass counterterm. The result of this computation is
\begin{equation}
    G_0 = \left( \begin{array}{ccc}
         \frac{i}{q^2} & 0 & 0\\
         0 & -\frac{1}{q}\tau_0 & \kappa_0\\
         0 & - \kappa_0& -q\tau_0  
    \end{array}\right) + \left(\begin{array}{ccc}
        0\quad &\quad0\quad &\quad0\quad   \\
        0\quad  & \quad0\quad & \quad0\quad \\
        0\quad &\quad0\quad &\text{div}
    \end{array}\right),
    \label{eq:G0}
\end{equation}
where, for $m>0$,\footnote{For $m<0$, $\kappa_0$ simply changes sign and $\tau_0$ remains the same.} 
\begin{equation}
\begin{aligned}
\kappa_0 & =-\frac{m}{2\pi q} \tanh ^{-1} \frac{q}{2m}, \\
\tau_0 & =-\frac{im}{4 \pi q}-\frac{i\left(4m^2+q^2\right)}{4 m q} \kappa_0,
\label{kappa tau}
\end{aligned}
\end{equation}
and $\text{div}$ has both a divergent and finite part that violate gauge invariance:
\begin{equation}
    \text{div} = \frac{i}{16 \pi}\left(\frac{\Lambda}{2}+(2+4 \log 2) m\right).
    \label{div}
\end{equation}

Although our $G_0$ — specifically its anti-Hermitian component, $G_0^A$ — contains a divergent term, it contributes to the norm of the eigenstates (\ref{norm}) only as a part of the following combination 
\begin{equation}
    V^{-1} + 2iG_0^A = 2 \left(\begin{array}{ccc}
      -\frac{1}{q^2} & 0   & -\frac{1}{e} \\
      0  & \frac{1}{e^2}+\frac{1}{q}\text{Im}(\tau_0) &  i\text{Re}(\kappa_0)\\
      -\frac{1}{e} & -i\text{Re}(\kappa_0)& i\cdot \text{div} - \frac{1}{e^2}m_A^2 + q \text{Im}(\tau_0)
    \end{array}\right).
\end{equation}
Hence, the following choice of the infinite part $m_A^{2\text{   }\infty}$ of the photon mass makes this combination finite:
\begin{equation}
m_A^2 = m_A^{2\text{   }\infty} + m_A^{2\text{ finite}}, \qquad    m_A^{2\text{   }\infty} = 
    ie^2\text{div}^\infty = 
    -\frac{e^2}{32\pi }\Lambda.
    \label{ma}
\end{equation}
where $\text{div}^\infty$ is the infinite part of div. Thus, we have derived the Hamiltonian eigenstates and fixed the infinite part of the photon mass counterterm such that the norms of these eigenstates are finite. We still need to determine the finite part of $m_A^2$, which we will do in the next section by computing the spectral density and comparing its perturbative expansion to the covariant calculation.

\section{Spectral Density}
\label{sec:correlator}

Our goal now is to compute the spectral density for the currents $j_\mu = \bar{\Psi} \gamma_\mu \Psi$ and the field strength tensor $F_{\mu\nu} = \partial_\mu A_\nu - \partial_\nu A_\mu$. It is sufficient to consider the spectral densities of $A_\perp, j_-$, and $j_\perp$.\footnote{The correlators of $j_+$ can be reconstructed from them using current conservation, which implies $j_+ = -\frac{1}{\partial^{+}} \left( \partial^-j_- + \partial^\perp j_\perp \right)$. As for $F_{\mu \nu}$, we are only interested in components with the physical photon mode $A_\perp$. Without loss of generality, we choose to consider $F_{\mu \perp}$, which becomes $F_{\mu \perp} = i p_\mu A_\perp \propto A_\perp$ in our frame $p_\mu = \left(\frac{p^2}{2}, 1, 0\right)$. This is also why $\langle A_\perp A_\perp \rangle $ is gauge invariant.\\
} At $q_+>2m^2$, the spectral density is given by the sum over the two eigenstates we found in the previous section:
\begin{align}
    \pi\rho_{ij}(q) &= \sum_{n,m = 0}^{2} \braket{\mc{O}_i(0)}{\psi_n,q}(\mc{N}^{-1})_{nm}\braket{\psi_m,q}{\mc{O}_j(0)},
    \label{5.12}
\end{align}
where $\ket{\psi_n,q}, \ket{\psi_m,q}$ are eigenstates of the Hamiltonian with momentum $q$, and $\mc{N}$ denotes the Gram matrix defined in (\ref{gram matrix}).

The components of $j_\mu$ are related to the free theory conserved current $j_\mu^{(0)}$ by 
\begin{equation}
\begin{gathered}
    j_- = \sqrt{2}\psi^*\psi = j_-^{(0)}, \\
    j_\perp = i(\psi^*\chi - \chi^*\psi) = j_\perp^{(0)} -\frac{1}{N_f} \frac{ie}{\sqrt{2}}\left(\psi \frac{1}{\partial_-} A_\perp\psi^* + \psi^* \frac{1}{\partial_-} A_\perp\psi \right).
    \end{gathered}
\end{equation}

However, the $j_\perp$ two-point function contains a divergent contribution, since the fermions in  the last expression are not normal ordered.  To regulate the contribution where they contract internally with each other, we introduce a new parameter $b$
\begin{equation}
\begin{aligned}
    \langle  j_\perp(x) j_\perp(0)\rangle  &=  \langle j_\perp^{(0)}(x) j_\perp^{(0)}(0)\rangle \\
     &- \frac{e^2}{2N_f^2} \left\langle  \left(\psi \frac{1}{\partial_-} A_\perp \psi^* + \psi^* \frac{1}{\partial_-} A_\perp \psi \right)(x)\left(\psi \frac{1}{\partial_-} A_\perp \psi^* + \psi^* \frac{1}{\partial_-} A_\perp \psi \right)(0) \right\rangle \\
    &= \langle j_\perp^{(0)}(x) j_\perp^{(0)}(0)\rangle + e^2 b^2 \langle A_\perp(x) A_\perp(0)\rangle,
    \label{74}
\end{aligned}
\end{equation}
up to subleading corrections in $1/N_f$. We will fix the value of $b$ shortly, but first notice that
\begin{equation}
    b \propto \contraction{}{A}{}{B} \psi^* \psi,
\end{equation}
so fixing $b$ corresponds to regulating the divergent part of the local operator $j_\perp$. With this definition, the currents are effectively given up to subleading in $1/N_f$ corrections by the following expressions 
\begin{align}
    j_- &= j_-^{(0)},& j_\perp &= j_\perp^{(0)} + e\cdot b A_\perp.
\end{align}

By rewriting these operators using the notation $\ket{\alpha_i}$ from the previous section, one obtains 
\begin{align}
    \mc{O}_i(x) &= \sum_{j=0}^{3}X_{ij}\alpha_j(x), &  X &= \left(\begin{array}{ccc} 
    1 & 0 & 0 \\
    0 & 1 & 0\\
    e b & 0 & 1
    \end{array}\right).
    \label{eq:Odef}
\end{align}

Now, in order to compute the spectral density for these states, we first need to choose the eigenbasis $\ket{\psi_n,q}$. We are allowed to select any 2 orthonormal eigenstates as (\ref{5.12}) is invariant under change of basis. According to the previous section, the eigenstates are parametrized by 5 parameters $S_i, C_i$ constrained by 3 equations (\ref{S=1}). Thus, for every state $\ket{\psi_n,q}$ we fix the values of 2 parameters according to our preference. Namely, for $\ket{\psi_1,q}$ we choose $C_1 = 1$ and $C_2 = 0$, and for $\ket{\psi_2,q}$ we choose $C_1 = 0, C_2 = 1$. Let us rewrite the $C_i$'s as a vector $C_{\psi_n} = \left(0, C_1, C_2\right)^T$ for convenience. Now, the Gram matrix from (\ref{norm}) takes the form 
\begin{equation}
     \mathcal{N} = 2\left(i + 2\left(V^{-1} + 2iG_0^A \right)^{-1} G_0^H\right)^\dagger G_0^H \left(i +2 \left(V^{-1} + 2iG_0^A \right)^{-1} G_0^H\right).
\end{equation}
Next, we need the overlaps $\braket{j_i(0)}{\psi_n}$ of the states in the spectral density with the Hamiltonian eigenbasis. These overlaps are given by
\begin{equation}
    \braket{\mc{O}(0)}{\psi_n,q} = X^* V^{-1} S_{\psi_n},
\end{equation}
where $\mc{O}$ denotes a vector with elements $\mc{O}_i$. 

Thus, equation (\ref{5.12}) takes the form
\begin{equation}
\begin{aligned}
   \pi \rho &= \sum_{n,m} X^* V^{-1} S_{\psi_n} \mc{N}^{-1}_{nm} S_{\psi_m}^\dagger (V^{\dagger})^{-1} X^T  \\
   &= X^*   \left( \left(1 + 2i G_0^A V\right)^\dagger(G_0^H)^{-1}\left(1 + 2iG_0^A V\right) + 4 V^\dagger G_0^H V\right)^{-1} X^T  \\
  & =X^*   \left( \left(1-2iG_0^\dagger V\right)^\dagger(G_0^H)^{-1}\left(1 - 2iG_0^\dagger V\right) \right)^{-1} X^T.
\end{aligned}    
\end{equation}
Therefore, the final expression we obtain for the spectral density is
\begin{equation}
     \pi \rho = X^*  \left(1- 2i G_0^\dagger V\right)^{-1}G_0^H\left(1 + 2i V G_0\right)^{-1}  X^T,
     \label{rho}
\end{equation}
where $\rho$ is a two-by-two matrix with elements $\rho_{ij}$ defined in (\ref{5.12}).

Now we aim to determine the value of the photon mass counterterm $m_A^2$ and of the parameter $b$ from the definition of $j_\perp$. To do so, we match our spectral density to the perturbative result, order by order up to $e^2$.  Since the theory is super-renormalizable, and no new divergences arise beyond $O(e^2)$, this procedure is sufficient to fix the counterterms exactly.

In the covariant approach, the standard computation of resumming the bubble Feynman diagrams to all orders in the coupling gives 
\begin{align}
    \langle j_\mu j_\nu \rangle &= \frac{q_\mu q_\nu - q^2 \eta^{\mu\nu}}{|q|} \left(\frac{q \left(-q \tau_0 +i e^2 \left(\kappa_0 ^2+\tau_0 ^2\right)\right)}{-e^4 \kappa_0 ^2+\left(q-i e^2 \tau_0 \right)^2}\right) + \varepsilon_{\mu\nu\rho}q^\rho \left( \frac{\kappa_0  q^2}{-e^4 \kappa_0 ^2+\left(q-i e^2 \tau_0 \right)^2}\right),
    \label{eq:Feynman}
\end{align}
with $\kappa_0, \tau_0$ defined in (\ref{kappa tau}). Adding the photon and putting all physical components together gives
\begin{equation}
    G_{\rm cov} = 
    \left(\begin{array}{ccc}
      \langle A_\perp A_\perp \rangle & \langle A_\perp j_- \rangle & \langle A_\perp j_\perp \rangle \\
       \langle j_- A_\perp \rangle  & \langle j_- j_- \rangle& \langle j_- j_\perp \rangle\\
        \langle j_\perp A_\perp \rangle  &\langle j_\perp j_-\rangle & \langle j_\perp j_\perp\rangle
    \end{array} \right)
    = G_{0,r}\left( 1 + 2i V_rG_{0,r}\right)^{-1},
    \label{eq:CovEqn}
\end{equation} \\
where $ G_{0,r}$ is the renormalized free theory two-point function, related to $G_0$ from (\ref{eq:G0}) by $G_0\vert_{\text{div} = 0}$, and $V_r = V\vert_{m_A = 0}$ is the renormalized potential of the theory.
Then, the desired spectral density is given by its Hermitian part
\begin{equation}
\begin{aligned}
    \pi\rho_{\rm cov} & = G_{\rm cov}^H = \frac{G_{\rm cov} + G_{\rm cov}^\dagger}{2} = \frac{G_{0,r}\left(1+2i V_r G_{0,r}\right)^{-1}+\left(1-2i G_{0,r}^{\dagger} V_r\right)^{-1} G_{0,r}^{\dagger}}{2}   \\
    & = \left(1 - 2i G_{0,r}^{\dagger} V_r\right)^{-1} G_{0,r}^H \left(1+2i V_r G_{0,r}\right)^{-1}.
    \label{rho diag}
    \end{aligned}
\end{equation}

Comparing equations (\ref{rho}) and (\ref{rho diag}) one can see that they have a quite similar form. Moreover, as $G_0^H = G_{0,r}^H$ does not depend on the coupling $e$, the condition $\rho = \rho_{\rm cov}$ is equivalent to 
\begin{equation}
    \left(1 + 2i V G_0\right)^{-1}  X^T = \left(1+2i V_r G_{0,r}\right)^{-1}.
    \label{eq:matching}
\end{equation}
Expanding in powers of $e$ and matching at first order, we find the following condition:
\begin{equation}
    -i q \tau_0 = b+i\left( \text{div} - q\tau_0\right),
\end{equation}
which gives 
\begin{equation}
    b = -i \cdot \text{div} =  \frac{1}{16 \pi}\left(\frac{\Lambda}{2}+(2+4 \log 2) m\right),
\end{equation}
 with $\rm div$ defined as in (\ref{div}). Next, matching the second order piece and plugging in $b = -i \cdot \text{div}$ gives the condition
\begin{equation}
    1+\frac{ie^2\tau_0}{q} = 1-\frac{ie^2\left(\text{div} - q\tau_0\right) - m_A^2}{q^2}.
\end{equation}
As one would expect, this condition for the photon mass  comes from $\langle A_\perp A_\perp \rangle$. Clearly, it gives 
\begin{equation}
    m_A^2 = i e^2\text{div} = -\frac{e^2}{16 \pi}\left(\frac{\Lambda}{2}+(2+4 \log 2) m\right),
\end{equation} 
 where now we have fixed the finite part of $m^2_A$ that was left undetermined in  (\ref{ma}).  With this choice, now the LHS and RHS of (\ref{eq:matching}) agree completely.

Note that we did not rely on the explicit form of $\kappa_0$ and $\tau_0$ given by (\ref{kappa tau}). Therefore, it is not too hard to see how our derivation should be applied to any super-renormalizable large $N$ vector-like theory, where the exact correlators satisfy an equation of the form (\ref{eq:CovEqn}). We discuss this generalization in more detail in Section \ref{sec:general}.

To summarize, in this section we have derived the spectral density for the conserved currents and the gauge field, and the result agrees with Feynman diagrams.

\section{One-Particle Eigenstates}
\label{sec:photon}

 Now, we are going to solve the Schr\"odinger equation for the one-particle states, i.e.~$p_+<2m$. As we have discussed, the hard momentum cutoff and the photon mass counterterm explicitly break gauge invariance, so we need to make sure that gauge invariance is preserved by their combination. In a gauge invariant theory in $d = 2+1$ with parity violation, a photon can have a non-zero physical mass $m_{ph}$ \cite{PhysRevD.33.3704}. A simple way to understand this is that when parity is violated, a Chern-Simons term is allowed, and then the photon gets what is commonly called a `topological mass' as can be seen by solving its equation of motion. In the limit of large fermion mass $m \gg e^2$, the physical photon mass is $m_{ph} = \frac{e^2}{4\pi}$. Let us show that we reproduce this result.

As before, the Schr\"odinger equation is 
\begin{equation}
    \left(p_+ - H_{\text{free}} \right) |\psi ,p \rangle =  V_{ij}|\alpha_i\rangle \langle \alpha_j |\psi,p \rangle.
    \label{sch again}
\end{equation} 
At energies below the two-fermion threshold, the kernel of $(p_+-H_{\rm free})$ no longer contains free two-fermion states.  Because $H_{\rm free} | A_\perp\rangle = 0$, one might think that we need to treat the cases $p_+=0$, where $|A_\perp\rangle$ is in the kernel of $(p_+-H_{\rm free})$, and $p_+>0$, where it is not, separately.  However, conveniently, it is not too hard to see that in either case the most general solution takes the form
\begin{equation}
    \ket{\psi,p} = \sum_{i=1}^{2}S_i \frac{P.V.}{p_+ - H_{\text{free}}}\ket{\alpha_i}+ C_0 \ket{\alpha_0},
    \label{eq:SinglePcleAnsatz}
\end{equation}
because if $p_+=0$, then the term $C_0 |\alpha_0\rangle$ takes care of the kernel of $(p_+ - H_{\rm free})$, and if $p_+>0$, then $C_0 |\alpha_0\rangle \equiv S_0 \frac{1}{p_+} |\alpha_0\rangle$ takes care of the missing $i=0$ term from the sum above.

Plugging (\ref{eq:SinglePcleAnsatz}) into (\ref{sch again}), and using that $H_{\text{free}}\ket{\alpha_0}=0$ gives the following equation: 
\begin{equation}
    \sum_{i=1}^{2}S_i \ket{\alpha_i}+ p_+C_0 \ket{\alpha_0} = -2i\sum_{k=1}^{2} \ket{\alpha_i} V_{ij} G_{0jk}^A S_k + C_0 V_{i0}\ket{\alpha_i} .
\end{equation}
Because the states $|\alpha_i\rangle$ are linearly independent, we can convert the equation above into three separate equations:
\begin{equation}
        \left\{ \begin{aligned}
        \frac{q^2}{2} C_0 &= -ie \text{Re}(\kappa_0) S_1 +ie\left(\text{div}- iq\text{Im}(\tau_0)\right) S_2+ C_0 \frac{m_A^2}{2} \\
        S_1 &= -\frac{e^2}{q}\text{Im}(\tau_0)S_1-ie^2 \text{Re}(\kappa_0)S_2 \\
        S_2 &= -\frac{e}{2}C_0
    \end{aligned} \right.,
\end{equation}
where $q= \sqrt{2p_+}$.
Eliminating $S_1, S_2$ we can see that the system only has solutions if the following equation on $q$ is satisfied 
\begin{equation}
    \left(e^2\text{Im}(\tau_0) + q\right)^2 -  e^4 \text{Re}(\kappa_0)^2 = 0,
    \label{q cond}
\end{equation}
where $\kappa_0$ and $\tau_0$ are given by (\ref{kappa tau}).  This condition coincides with the condition for the denominator of the correlation function (\ref{eq:Feynman}), derived via Feynman diagrams, to vanish (recall that at $q<2m$, $\tau_0$ and $\kappa_0$ are pure imaginary and pure real, respectively). 

Equation (\ref{q cond}) is impossible to solve analytically, though straightforward to solve numerically. It can also be solved perturbatively in an expansion in small $e^2/m$, where one can see that it has a solution at $q \sim e^2$, corresponding to a massive photon. More precisely, expanding $\tau_0$ and $\kappa_0$ up to linear order at $q \rightarrow 0$ gives $\text{Im}(\tau_0) = \frac{q}{12\pi m}$  and $\text{Re}(\kappa_0) = -\frac{1}{4\pi}$. Then (\ref{q cond}) has the following solution at $e^2 \ll m$:
\begin{equation}
    q = \frac{e^2}{4\pi}.
\end{equation}
This is the physical mass of the photon for $e^2 \ll m$. One can independently derive it in this limit by integrating out the heavy fermions and seeing that they generate a Chern-Simons term for the $U(1)$ gauge field. One of course obtains the same result this way.
Beyond the limit of small $e$, one can simply plot the LHS of (\ref{q cond}) as a function of $q^2$, and see that for $0< e^2 < e_*^2=16\pi m$ it has a solution at real $q<2m$, corresponding to the physical massive photon.  

In the class of models we have considered, there is no positron/electon bound state below the $2m$ threshold, despite the fact that the photon produces an attractive force between the two particles. This is roughly due to the fact that when the photon is massless, it produces a confining $\log r$ potential between the particles which can produce a relativistic bound state even at small coupling.   More generally, by modifying the model it is possible to also have additional single-particle states beyond just the massive photon.  A simple example is to add a bare Chern-Simons term $\frac{k e^2}{4\pi} A d A$ to the original Lagrangian, which has the effect of shifting $\kappa_0$ by $-\frac{k}{2\pi}$. The photon mass cuts off the $\log(r)$ growth in the electromagnetic potential, and if the photon mass is sufficiently large then at small $e$ the nonrelativistic approximation implies that there should be a bound state slightly below the $2m$ threshold.  Expanding (\ref{q cond}) around $q\approx 2m$,
\begin{equation}
    q = m (2- \epsilon),
\end{equation}
there is a solution at
\begin{equation}
    \log \epsilon = \frac{4\pi m}{e^2} + k + \log 4 - 1/4
\end{equation}
To be a nonrelativistic bound state, the binding energy must be small compared to the mass, which requires $\log \epsilon < 0$, and therefore  only exists if $k$ is below $-\frac{4 \pi m}{e^2} + O(e^0)$. 

\section{S-Matrix and Form Factors}
\label{sec:smatrix}

We now want to derive the S-matrix for large $N$ vector-like theories by constructing scattering states with the Lippmann-Schwinger equation:
\begin{align}
    \ket{\Psi^\pm, p} &= \ket{\psi_\text{free},p} + (p_+ - H_\text{free} \pm i0)^{-1}V\ket{\Psi^\pm,p},
\end{align}
where $\ket{\Psi^\pm}$ are the ``in'' and ``out'' states respectively.\footnote{In \cite{Gabai:2022snc}, the authors also solved the Lippmann-Schwinger equation in a large $N$ limit for a 3d gauge theory in lightcone quantization.  However, the model they considered was that of a nonabelian Chern-Simons $U(N)$ gauge theory in the large $N$ limit, where as usual the topology of diagrams that must be resummed is significantly more complicated. Specifically, our calculation simplifies after equation (\ref{eq:scatteringcoeffs}) because the interactions have the factorized form of large $N$ vector-like theories.   } Note that this representation of the scattering states is extremely similar to our formula (\ref{general sol}) for the Hamiltonian eigenstates, suggesting that we should be able to apply similar methods as before but now to the calculation of scattering amplitudes. For simplicity, we will just focus on the singlet asymptotic states, and keep the leading large $N_f$ piece.  The full S-matrix has the flavor structure
   \begin{align*}
    \mc{M}^{ijk\ell} &= \left(\mc{M}(s)\delta^{ij}\delta^{k\ell} + \mc{M}(t)\delta^{ik}\delta^{j\ell} + \mc{M}(u)\delta^{i\ell}\delta^{jk}\right),
\end{align*}
and we only compute the coefficient of $\delta^{ij} \delta^{k\ell}$.  

First, recall that our $V$ may be decomposed as 
\begin{align}
    V &= V_{ij}\ket{\alpha_i}\bra{\alpha_j} .
\end{align}
Inserting $V$ into the Lippmann-Schwinger equation yields
\begin{align}
    \ket{\Psi^\pm, p} &= \ket{\psi_\text{free},p} + \frac{1}{p_+ - H_\text{free} \pm i0 }\ket{\alpha_i}\Big[V_{ij}\braket{\alpha_j}{\Psi^\pm,p}\Big]. 
\end{align}
If we now define coefficients 
\begin{align}
    S_i^\pm &\equiv V_{ij}\braket{\alpha_j}{\Psi^\pm,p}, \label{eq:scatteringcoeffs}
\end{align}
our equation for the scattering states may be rewritten as 
\begin{align}
    \ket{\Psi^\pm, p} &= \ket{\psi_\text{free},p} + \frac{1}{p_+ - H_\text{free}\pm i 0} \ket{\alpha_i}S_i^\pm.
\end{align}
Again, this form of the equation now closely matches what we obtained in Section \ref{sec:eigenstates}, and can be solved in a similar manner. We solve for the coefficients $S_i^\pm$ using (\ref{eq:scatteringcoeffs}), yielding 
\begin{align}
    S_i^+ 
    &= \left[(V^{-1} + 2iG_0)^{-1}\right]_{ik}\langle \alpha_k | \psi_\text{free}, p \rangle, \\
     S_i^-
     &= \left[(V^{-1} - 2iG_0^\dagger)^{-1}\right]_{ik}\langle \alpha_k | \psi_\text{free},p \rangle,
\end{align}
where we recall our definition of $G_{0,ij}$
\begin{equation}
\begin{aligned}
    2iG_{0,ij} = -\langle \alpha_i | \frac{1}{p_+ - H_{\rm free} + i \epsilon} |\alpha_j \rangle 
\end{aligned}
\end{equation}
as a three-by-three matrix.  To avoid clutter, we can write the equations for $S_i^\pm$ in the following index-free notation:
\begin{align}
    S^+ &= A \cdot \langle \alpha |\psi_{\rm free} ,p \rangle, &
     S^- &= A^\dagger \cdot \langle \alpha |\psi_{\rm free},p \rangle, 
     \label{eq:Adef}
\end{align}
with the matrix $A$ defined as 
\begin{align}
    A & \equiv (V^{-1}+2iG_0)^{-1} = (V_r^{-1} + 2 i G_{0,r})^{-1}.
    \label{eq:Adefn}
\end{align}

As a check, we can compute the norm of the scattering states as follows:
\begin{equation}
    \begin{aligned}
        \langle \Psi^+ , q | \Psi^+ , p \rangle &= \langle \psi_{\rm free}, q | \psi_{\rm free} , p \rangle \\
        &+ \langle \psi_{\rm free} ,q| \frac{1}{p_+ - H_{\rm free} + i 0 } |\alpha_i \rangle S^+_i + (S_i^+)^* \langle \alpha_i | \frac{1}{q_+ - H_{\rm free} - i 0 } |\psi_{\rm free}, p \rangle \\
        & + (S_i^+)^* \langle \alpha_i | \frac{1}{q_+ - H_{\rm free} -i0} \frac{1}{p_+-H_{\rm free} +i0} | \alpha_j\rangle S_j^+. \label{eq:overlap_check}
    \end{aligned}
\end{equation}
We can simplify the expression above using the fact that $|\psi_{\rm free},p\rangle$ is an eigenstate of $H_{\rm free}$ with eigenvalue $p_+$. We will also only keep track of the $\delta(p_+-q_+)$ contributions since they are the only pieces that contribute to the norm, so we can drop the principal value pieces $\frac{P.V.}{p_+ - q_+}$. Then, terms in the second line of (\ref{eq:overlap_check}) simplify to
\begin{equation}
\begin{aligned}
    \langle \psi_{\rm free} ,q | \frac{1}{p_+ - H_{\rm free} + i 0 } |\alpha_i \rangle S_i^+ &= \frac{1}{p_+-q_++i0} \langle \psi_{\rm free}, q| \alpha_i \rangle S_i^+\\
    &= -i \pi \delta(p_+ - q_+) \langle \psi_{\rm free},q| \alpha\rangle A \langle \alpha | \psi_{\rm free}, p\rangle .
    \end{aligned}
\end{equation}
Finally, we can also use the identity (\ref{eq:DtildeDtildeDhat}) to simplify the third line of (\ref{eq:overlap_check}) to
\begin{equation}
    \begin{aligned}
       & (S_i^+)^* \langle \alpha_i | \frac{1}{q_+ - H_{\rm free} -i0} \frac{1}{p_+-H_{\rm free} +i0} | \alpha_j\rangle S_j^+ \\
       & = \pi \delta(q_+-p_+) \langle \psi_{\rm free},q | \alpha\rangle \left( A^\dagger (4 G_0^H) A  \right) \langle \alpha | \psi_{\rm free} , p \rangle.
       \label{eq:ThirdLine}
    \end{aligned}
\end{equation}
Putting this all together, the norm is given by
\begin{equation}
    \begin{aligned}
        \langle \Psi^+ , q | \Psi^+ , p \rangle &= \langle \psi_{\rm free}, q | \psi_{\rm free} , p \rangle \\
        & + \pi \delta(q_+ - p_+)  \langle \psi_{\rm free},q | \alpha\rangle \left( -i A + i A^\dagger + A^\dagger (4 G_0^H) A \right) \langle \alpha | \psi_{\rm free}, p\rangle
    \end{aligned}
\end{equation}
The second line vanishes, as can be seen by multiplying on the left by $(A^\dagger)^{-1}$ and the right by $A^{-1}$ and using the fact that the anti-Hermitian piece of $A^{-1}$ is $2 i G_0^H$.

We are now ready to compute the S-matrix by taking the overlap $\braket{\Psi^-,q}{\Psi^+,p}$, where explicit momentum dependence will be suppressed from here on. The calculation proceeds along almost exactly the same lines as the calculation of the norm, since we simply flip the sign of some of the ``$i0$''s.  Now the analogous version of equation (\ref{eq:ThirdLine}) vanishes, and the remaining terms are
\begin{align}
    \langle \Psi^-,q | \Psi^+ ,p \rangle &= \langle \psi_\text{free} ,q| \psi_\text{free} ,p \rangle + i\mc{M}\times\pi\delta(q_+-p_+),
\end{align}
where we've defined 
\begin{equation}
\begin{aligned}
    \mc{M} & = -\braket{\psi_\text{free},q}{\alpha}\cdot\Big[A + A\Big]\cdot\braket{\alpha}{\psi_\text{free},p} \\
    &= -\braket{\psi_\text{free},q}{\alpha}\cdot\Big[2(V_r^{-1}+2iG_{0,r})^{-1}\Big]\cdot\braket{\alpha}{\psi_\text{free},p}.
    \label{eq:LargeNSmat}
\end{aligned}
\end{equation}

There is also an implicit spatial momentum-conserving $\delta$-function.  $\mc{M}$ is our final result for the scattering amplitude in the $s$-channel. Overlaps between the free theory two-particle Fock space state $\ket{\psi_\text{free}}$ and our basis states $\ket{\alpha}$ can be computed using a standard mode expansion.\footnote{To apply (\ref{eq:LargeNSmat}) to QED$_3$, for the fermionic fields $\psi_i(x)$ we can use the convention found in equation (A.3) in \cite{Delacr_taz_2019}, and $A_\perp(x) = \int \frac{d^2p}{(2\pi)^2}\frac{1}{\sqrt{2p_-}}e^{ip\cdot x}\left(\alpha_{-p}\Theta(-p_-) + \alpha_p^\dagger\Theta(p_-)\right)$ for the photon.} In appendix \ref{sec:O(N)SMatrix}, we apply the general expression (\ref{eq:LargeNSmat}) to the specific case of the 3d $O(N)$ model at infinite $N$.

Note that a side product of the construction of the scattering states, we also obtain the form factors for local operators. Specifically, by inspection of (\ref{eq:scatteringcoeffs}) and (\ref{eq:Adef}), the two-particle form factors of the local operators $\alpha_i$ are
 \begin{equation}
 \langle \alpha |\Psi^+,p \rangle = V^{-1} S^+ = (1 + 2 i G_0 V)^{-1} \langle \alpha | \psi_{\rm free} ,p \rangle ,
 \end{equation}
and similarly for $\langle \alpha |\Psi^- ,p\rangle$. However, recall that we need to renormalize the local operators, as in (\ref{eq:Odef}). Thus, the physical form factors are
 \begin{equation}
     \langle {\cal O} | \Psi^+,p\rangle = X^* V^{-1} (V_r^{-1}+2i G_{0,r})^{-1} \langle \alpha | \psi_{\rm free} ,p\rangle
     = (1+2i  G_{0,r}V_r)^{-1} \langle \alpha | \psi_{\rm free} ,p\rangle,
 \end{equation}
where we have used (\ref{eq:Adefn}) and $X^* V^{-1}= V_r^{-1}$.

\section{General Large $N$ Vector-Like Theory}
\label{sec:general}

By looking back at the Hamiltonian of the form (\ref{general H}) it is clear that it does not change much if the sum there is taken over any finite number of states $k$, rather than just the three states $|\alpha_i \rangle$. In fact,  we did not rely on the fact that our operator space is three-dimensional, or that the specific fields and Lagrangian were QED$_3$.  The only crucial ingredients were that the Hamiltonian coupled a finite set of single and two-particle states, and did not produce states with higher particle number, which is the structure of a general large $N$ vector-like theory. Thus our approach can be applied to a more general class of theories. In this section, we formulate the generalization and list the results for its eigenstates, spectral density, and S-matrix.

Let us consider an action with a scalar boson singlet $\phi$  and $m$ vector-like pairs $v_a, v_a^\dagger$, that transform in the fundamental representation of $SU(N)$: 
\begin{equation}
    v'_{a,\alpha} = \mathcal{O}\indices{_a^b}v_{b,\alpha} , \quad \mathcal{O}\indices{_a^b} \in SU(N).
\end{equation}
The index $\alpha$ is separate from the $SU(N)$ structure, and in the case of QED$_3$ they were spinor indices.
As before, we will consider the large $N$ limit of the theory. We will separate the mass parameter of the singlet $\phi$ into a renormalized mass and counterterm:
\begin{equation}
    m_\phi = m_{\phi, r} + \delta m_\phi.
\end{equation}
In order to have a stable $\phi$ particle, we will assume that $m_{\phi,r}$ is below the threshold to produce two of the vector-like particles. We also focus for simplicity on an action containing just one singlet, though extending the analysis to include additional singlets is straightforward. The most general formulation preserving particle number conservation, provided that the scalar field  is counted equivalently to two particles $v_a$, is given by the following Lagrangian: 
\begin{equation}
    \mathcal{L} = \mathcal{L}_{\text{free}} - \delta m_\phi \phi^2 - 2 \lambda_{i} \phi \left(v_a^\dagger T^i v_a\right) -g_{ij} \left( v_a^\dagger T^i v_a \right)\left( v_b^\dagger T^j v_b \right).
    \label{gen L}
\end{equation}

We have left the $\alpha$ indices implicit in this expression.  That is, $\left(v_a^\dagger T^i v_a\right)$ is shorthand for
\begin{equation}
    \left(v_a^\dagger T^i v_a\right) \equiv \left(v_{a,\alpha}^\dagger T^i_{\alpha \beta} v_{a,\beta}\right).
\end{equation}
To maintain generality, we leave the operators $T^i$ unspecified. For instance, in the context of a gauge theory, the operators $T^i$ correspond directly to $\gamma$ matrices.

Nearly identical arguments to those we used in the case of QED$_3$ lead to the following Hamiltonian density in the large $N$ limit:
\begin{equation}
    H = H_{\text{free}} + \delta m_\phi |\phi \rangle \langle \phi| + \lambda_{i} \left(|\phi \rangle \langle v_a^\dagger T^i v_a |+h.c.\right) + g_{ij} | v_a^\dagger T^i v_a \rangle \langle v_b^\dagger T^j v_b |
\end{equation}
where we use same notation as before 
 for $|\mathcal{O} \rangle = \mathcal{O}(0) |0\rangle$,  $H_{\text{free}}$ is the free theory Hamiltonian, and there is an implicit sum over $i, j = 1, ...,  k-1.$ 
Following (\ref{def V}) we introduce a similar notation $|\alpha_0\rangle = | \phi \rangle$, $|\alpha_i \rangle = | v_a^\dagger T^i v_a \rangle $, and $V_{00} = \delta m_\phi$, $V_{0i} = V_{i0} = \lambda_i$,  $V_{ij} = g_{ij}$. Equivalently, 
   \begin{equation}
       V = 
    \left(\begin{array}{c|cc}
    \delta m_\phi && {\lambda_i} \\ 
    \hline
    \lambda_i && {g_{ij}} 
    \end{array}\right). 
    \end{equation}
Then the Hamiltonian can be written as
\begin{equation}
    H = H_{\text{free}} +  V_{ij}|\alpha_i\rangle \langle \alpha_j |
    \label{general H2}
\end{equation}
with an implicit sum over $i, j = 0, 1, ..., k-1$. Now  that the Hamiltonian has the same form as the Hamiltonian for QED$_3$ (\ref{general H}), the same derivations apply. We will list the results for this theory without repeating the derivations, since they proceed along exactly the same lines as for the QED case.

For the eigenstates with energies above the mass of a vector-like pair, we get 
\begin{equation}
    |\psi_n,p \rangle  =  \left( 2 \frac{P.V. }{p_+ -  H_{\text{free}}} \left(V^{-1} + 2iG_0^A \right)^{-1}G_0^H  + \pi\delta\left( p_+ - H_{\text{free}}  \right) \right)C_{\psi_n} |\alpha\rangle,
\end{equation}
where $C_{\psi_n} = \left(0, C_1, C_2, ..., C_{k-1}\right)^T$. The overlaps of the eigenstates are given by
\begin{equation}
\begin{gathered}
\braket{\psi_n,q}{\psi_m,q'} = \mc{N}_{nm}2\pi \delta(q^2 - q'^2), 
\end{gathered}
\end{equation}
where the Gram matrix is
\begin{align}
    \mc{N}_{nm} &= 2C_{\psi_n}^\dagger\left(i + 2\left(V^{-1} + 2iG_0^A \right)^{-1} G_0^H\right)^\dagger G_0^H \left(i + 2\left(V^{-1} + 2iG_0^A \right)^{-1} G_0^H\right)C_{\psi_m}.
\end{align}

The spectral density of two $\alpha$-states takes the form
\begin{equation}
\begin{gathered}
   \pi \rho_{\alpha, \alpha} = 
     \left(1+ 2i G_0^\dagger V\right)^{-1}G_0^H\left(1 - 2i V G_0\right)^{-1}.
   \label{gen rho}
\end{gathered}    
\end{equation}

One can straightforwardly obtain the spectral density for any other basis of operators $\ket{\Phi_i}$ in the subspace spanned by $\ket{\alpha_i}$'s:
\begin{align}
    \ket{\Phi_i} &= \sum_{j=0}^{k}X_{ij}\ket{\alpha_j} \Rightarrow   \rho(\Phi_i, \Phi_j) = X_{ik}\rho_{\alpha_k, \alpha_l}X_{lj} .
\end{align}

The point of equation (\ref{gen rho}) is of course that it is what one would get by resummation of Feynman diagrams. This comparison is straightforward for a Lagrangian in the form we have assumed in equation (\ref{gen L}), where the $k \times k$ matrix of time-ordered two-point functions of the $\alpha_i$ operators from resumming Feynman diagrams
\begin{equation}
    G_{\rm cov} = G_{0,r} + G_{0,r} \left(-2iV_r\right) G_{0,r} + \dots =  G_{0,r} \left( 1 + 2iV_r G_{0,r}\right)^{-1}
\end{equation}
has a  Hermitian part of the same form as (\ref{gen rho}).

Once everything is expressed in terms of general matrices $G_0$ and $V$, one can also immediately apply the results in Section \ref{sec:smatrix} for the S-matrix, which by inspection made no reference to the explicit form of these matrices. Therefore, by solving the Lippmann-Schwinger equation, we obtain the same result for the leading large $N$ scattering amplitude in the flavor singlet channel:\footnote{We explicitly apply this formula to the $O(N)$ model at large $N$ in Appendix \ref{sec:O(N)SMatrix}.}
\begin{align}
    \mc{M}(s) &= -2\braket{\psi_\text{free},p}{\alpha_i}\left(V^{-1}+2iG_0\right)^{-1}_{ij}\braket{\alpha_j}{\psi_\text{free},p}.
\end{align}

\section{Conclusion and Outlook}

Although the infinite $N_f$ limit of QED$_3$ is solvable by traditional methods, as we have seen it is not immediately obvious how to take advantage of those simplifications in a Hamiltonian framework.  Ultimately, our aim in understanding how to make those simplifications manifest using Hamiltonians is to provide guidance on how to go to higher orders in $1/N_f$ and eventually to finite $N_f$.  

Starting at the next subleading order in $1/N_f$, a number of interesting complications arise, as particle number can change and therefore one must include mixing between the lowest particle number sector and the next-to-lowest particle sector.  As a result, the matrix elements of the interaction Hamiltonian will not just have the simple factorized form of (\ref{H QED}).

Nevertheless, one of the main simplifications of our approach has been that by choosing a basis of states dictated by the interactions themselves, one can find an accurate description of the energy eigenstates using a much smaller basis. Moreover, the fact that  Lippmann-Schwinger applied to this system successfully reproduced the S-matrix depended on maintaining certain analytic properties of the wavefunctions of the states, and suggest that perhaps at finite $N_f$ one could still use a basis of wavefunctions with the necessary properties even while diagonalizing the Hamiltonian numerically.

\acknowledgments

We thank Richard Brower and Ami Katz for helpful discussions. ALF, AN, and NR are supported by the US Department of Energy Office of Science under Award Number DE-SC0015845.

\appendix

\section{Appendix: $O(N)$ S-Matrix}
\label{sec:O(N)SMatrix}

As an application of our S-matrix machinery in Section \ref{sec:smatrix}, we can consider the much simpler $O(N)$ model at large $N$ and compute the singlet sector scattering amplitude. 

To begin, the $O(N)$ Lagrangian is given by
\begin{align}
    \mc{L} = \frac{1}{2}(\partial_\mu \phi_i)(\partial^\mu \phi_i) - \frac{1}{2}m^2\phi_i\phi_i - \frac{\lambda}{4N}(\phi_i\phi_i)(\phi_j\phi_j).
\end{align}
At infinite $N$, the lightcone Hamiltonian takes the form 
\begin{align}
    H &= H_\text{free} + \frac{\lambda}{4}| \phi^2 \rangle \langle \phi^2 |,
\end{align}
where we have defined 
\begin{align}
    | \phi^2 \rangle &\equiv \frac{1}{\sqrt{N}}\phi_i^2(0) | \text{vac} \rangle.
\end{align}
Thus, in the case of $O(N)$ we only have a one-dimensional subspace to consider.

Recall the general expression for the scattering amplitude 
\begin{align}
    \mc{M} &= -\langle \psi_\text{free} | \alpha \rangle \cdot \Big[2(V^{-1}+2iG_0)^{-1}\Big]\cdot\langle \alpha | \psi_\text{free}\rangle.
\end{align}
Taking $| \psi_\text{free} \rangle = | p_1^i, p_2^i\rangle$ to be a singlet state and $G_0$ to be the usual free theory correlator of $\phi^2$ operators, we find 
\begin{align}
    \mc{M}(s) &= -\frac{2\lambda}{1 + \frac{\lambda}{8\pi\sqrt{s}}\left[\log(\frac{\sqrt{s}+2m}{\sqrt{s}-2m}) + i\pi\right]}.
\end{align}
This is consistent with what was found in \cite{Henning_2023}.



\bibliographystyle{JHEP}
\bibliography{biblio.bib}

\end{document}